\documentclass{article}
\usepackage{spconf,amsmath,graphicx}

\usepackage{xcolor}
\usepackage{multirow}
\usepackage{float}
\usepackage{cite}
\usepackage{url}
\usepackage{amssymb}
\usepackage{algorithm}  
\usepackage{algpseudocode}  
\usepackage{pifont}
\newcommand{\cmark}{\ding{51}}%
\newcommand{\xmark}{\ding{55}}%

\usepackage{booktabs,multirow}
\usepackage{adjustbox}

\title{Enhancing Speech-to-Speech Translation with Multiple TTS Targets}
\name{Jiatong Shi$^{1}$, Yun Tang$^{2}$, Ann Lee$^{2}$, Hirofumi Inaguma$^{2}$, Changhan Wang$^{2}$, Juan Pino$^{2}$, Shinji Watanabe$^{1}$}
\address{
    $^{1}$ Carnegie Mellon University \quad $^{2}$ Meta AI\\
    \small{\texttt{\{jiatongs,swatanbe\}@cs.cmu.edu}, \texttt{\{yuntang,annl\}@meta.com}}
}

\begin{document}
\ninept
\maketitle
\begin{abstract}
It has been known that direct speech-to-speech translation (S2ST) models usually suffer from the data scarcity issue because of the limited existing parallel materials for both source and target speech. Therefore to train a direct S2ST system, previous works usually utilize text-to-speech (TTS) systems to generate samples in the target language by augmenting the data from speech-to-text translation (S2TT). However, there is a limited investigation into how the synthesized target speech would affect the S2ST models. In this work, we analyze the effect of changing synthesized target speech for direct S2ST models. We find that simply combining the target speech from different TTS systems can potentially improve the S2ST performances. Following that, we also propose a multi-task framework that jointly optimizes the S2ST system with multiple targets from different TTS systems. Extensive experiments demonstrate that our proposed framework achieves consistent improvements (2.8 BLEU) over the baselines on the Fisher Spanish-English dataset. 
\end{abstract}

\section{Introduction}
\label{sec: intro}

Speech-to-speech translation (S2ST) focuses on translating speech from a source language into the speech of a target language \cite{vidal1997finite}. Conventional cascaded S2ST models decompose the task into three components, including automatic speech recognition (ASR), machine translation (MT), and text-to-speech (TTS) \cite{matsuda2013multilingual, do2016preserving}. Alternatively, some previous works adopt end-to-end speech-to-text translation (S2TT) instead of ASR and MT. However, it would introduce high computational costs and inference latency for further application. To mitigate the issue, recent literature focuses on building direct S2ST models without three standalone modules \cite{jia2019direct, jia2022translatotron, jia2022leveraging, lee2022direct, kano2021transformer, zhang2021uwspeech, ma2021direct, popuri2022enhanced, lee2021textless}.

The training of direct S2ST models needs inevitably large amounts of parallel S2ST corpora, which are far more difficult to obtain than conventional cascaded methods \cite{jia2022leveraging}. To mitigate the issue and enable the training for S2ST models, previous works incorporated TTS systems to form the dataset for S2ST~\cite{jia2019direct, jia2022translatotron, jia2022leveraging, lee2022direct, kano2021transformer, zhang2021uwspeech, ma2021direct, popuri2022enhanced}. Nearly all the published datasets on S2ST are extended from speech-to-text corpora where the target speech for S2ST is synthesized by  TTS systems \cite{kikui2003creating, jia2022cvss, jeuris2022libris2s}. When synthesizing the target speech for S2ST, researchers in previous works usually select a specific TTS system. For instances, in \cite{jia2022cvss}, they utilized a variant of Non-attentive Tacotron (NAT) \cite{shen2020non}, while in \cite{jeuris2022libris2s}, they adopted Fastspeech2 \cite{ren2020fastspeech}. To the best of our knowledge, there is no investigation into how different synthesized target speech would affect the S2ST modeling.

To fill the research gap aforementioned, this paper focuses on the effect of different synthesized speech from various TTS systems. We find simply using training data from multiple TTS systems can improve the performance of S2ST. To further utilize the shared knowledge across multiple TTS systems, we further propose a framework that jointly optimized the S2ST systems with multiple targets from different TTS systems. Results show that our proposed method could significantly improve the S2ST performances over baseline models. To be specific, our proposed framework shows a 2.8 BLEU score improvement over the best baseline system with a single TTS target on the Fisher Spanish-English dataset \cite{post2014fisher}. The contribution of this work can be summarized as follows:
\begin{itemize}
    \item We first investigate the effect of different TTS systems for target synthesized speech for S2ST.
    \item We propose a multi-task framework that combines knowledge from different TTS data, which shows reasonable improvements according to our experiments.
\end{itemize}


\section{Methodology}

In this section, we first review the background of this research, including the S2ST system with discrete units and various TTS systems used in this work. Then, we introduce our proposed framework for combining knowledge from target speech from different TTS systems.

\subsection{Background}
\label{ssec: background}

\noindent \textbf{S2ST with discrete units}: Speech self-supervised learning (SSL) models have shown outstanding performances on various tasks \cite{yang2021superb, tsai2022superb}. Notably, they are also applicable to synthesis tasks \cite{polyak2021speech, huang2022s3prl, hayashi2020discretalk}. To apply SSL representations, a common strategy is to discretize them into speech units through clustering approaches \cite{polyak2021speech, lakhotia2021generative}. Previous works have shown that the discrete units can disentangle linguistic content from other acoustic properties (e.g., speaker identity or prosody information), resulting in easier learning of linguistic information directly from speech \cite{polyak2021speech}. Due to this reason, Lee et al. proposed a direct S2ST model, which uses discrete speech units as the prediction target of the system \cite{lee2022direct}. Their experiments also demonstrated their superiority over the translatotron-based methods \cite{jia2019direct, jia2022translatotron} and comparable performances to the cascaded S2ST systems \cite{aguero2006prosody, matsuda2013multilingual, do2016preserving}. The discrete units in their system are generated from the K-Means clustering over the representation from a pre-trained HuBERT model \cite{hsu2021hubert}. 



\noindent \textbf{TTS systems}: As mentioned in Sec.~\ref{sec: intro}, previous studies usually employ TTS systems to generate the target speech for S2ST. The translatotron series of works mainly adopt auto-regressive (AR) TTS systems (i.e., NAT) \cite{jia2019direct, jia2022cvss, jia2022translatotron}, while there are also other studies that apply non-auto-regressive (NAR) TTS such as Fastspeech2 \cite{jeuris2022libris2s}. All these models are text2Mel models, where they convert the text to Mel spectrogram, so they need additional vocoders to get the waveform of speech. The choices of vocoders also vary, including non-parametric Griffin-Lim and neural vocoders.

In this work, we select three TTS models: Tacotron2 (TT2) \cite{shen2018natural}, Fastspeech2 (FS2) \cite{ren2020fastspeech}, and VITS \cite{kim2021conditional}. Tacotron2 is a classical AR TTS text2Mel model, while Fastspeech2 is a typical NAR TTS text2Mel model. VITS, different from others (text2Mel + vocoder), directly models the process from text to waveform (text2wav), which does not need additional vocoders. For text2Mel models (i.e., TT2 and FS2), we adopt three different vocoders for investigation: Parallel WaveGAN (PWG) \cite{yamamoto2020parallel}, Hifi-GAN (HFG)~\cite{kong2020hifi}, and StyleMelGAN~(SMG)~\cite{mustafa2021stylemelgan}. We also investigate the effect of duration control for NAR models by tuning the speed factor in the inference.

For the first set of our experiments, we evaluate the S2ST model with different target TTS speech. Then, we further conduct experiments on combining synthesized speech from different TTS systems. For this study, we focus only on single-speaker TTS systems.

\noindent \textbf{Overall Workflow}: Based on the previous work discussed, the overall workflow for constructing an S2ST system is outlined below: (1) Target speech synthesis: Target speech is synthesized using a TTS model, which can be either an acoustic model and vocoder or a direct text2wav model. (2) Discrete unit extraction: The synthesized target speech is converted into discrete units using a HuBERT model by clustering. (3) S2ST system training: The S2ST model is trained using source speech as input and target discrete units as output. (4)~Inference: During inference, the S2ST model converts the source speech into a sequence of discrete units. Then, a unit-based vocoder is applied to generate the final waveform speech.

\begin{figure}[t]
  \centering
  \includegraphics[width=0.9\linewidth]{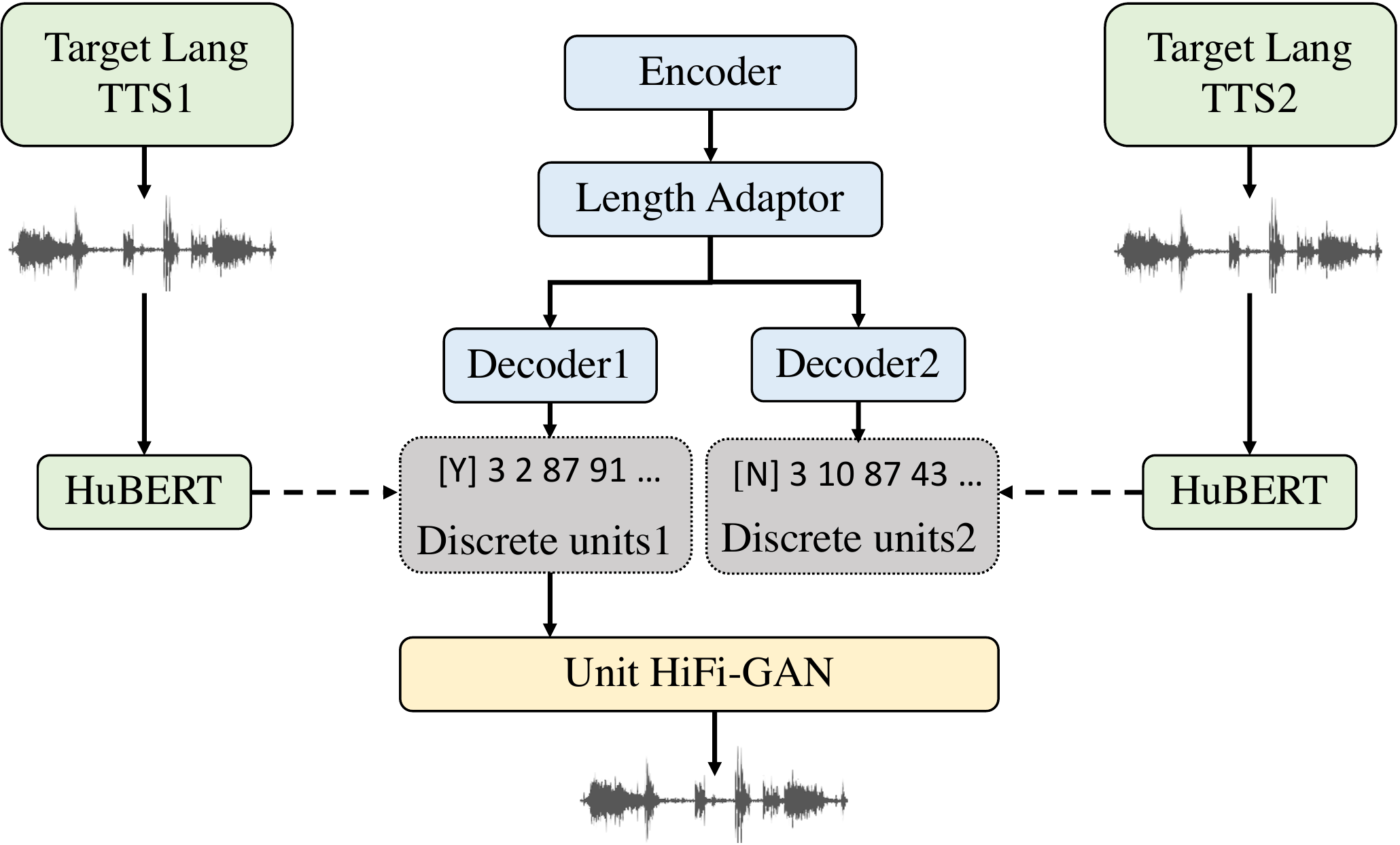}
  \vspace{-10pt}
  \caption{The framework of our proposed 
S2ST model using multiple TTS targets. The blue blocks represent the S2ST modeling; the green blocks are modules used to generate target discrete units; the gray blocks are the targets of the S2ST model, while the first token is for predicting which TTS target is used for inference; the yellow block is for model inference. Details are explained in Sec.~\ref{ssec: proposed framework}.}
  \label{fig:framework}
  \vspace{-15pt}
\end{figure}

\subsection{The Proposed Framework}
\label{ssec: proposed framework}

\begin{figure}[t]
  \centering
  \includegraphics[width=0.5\linewidth]{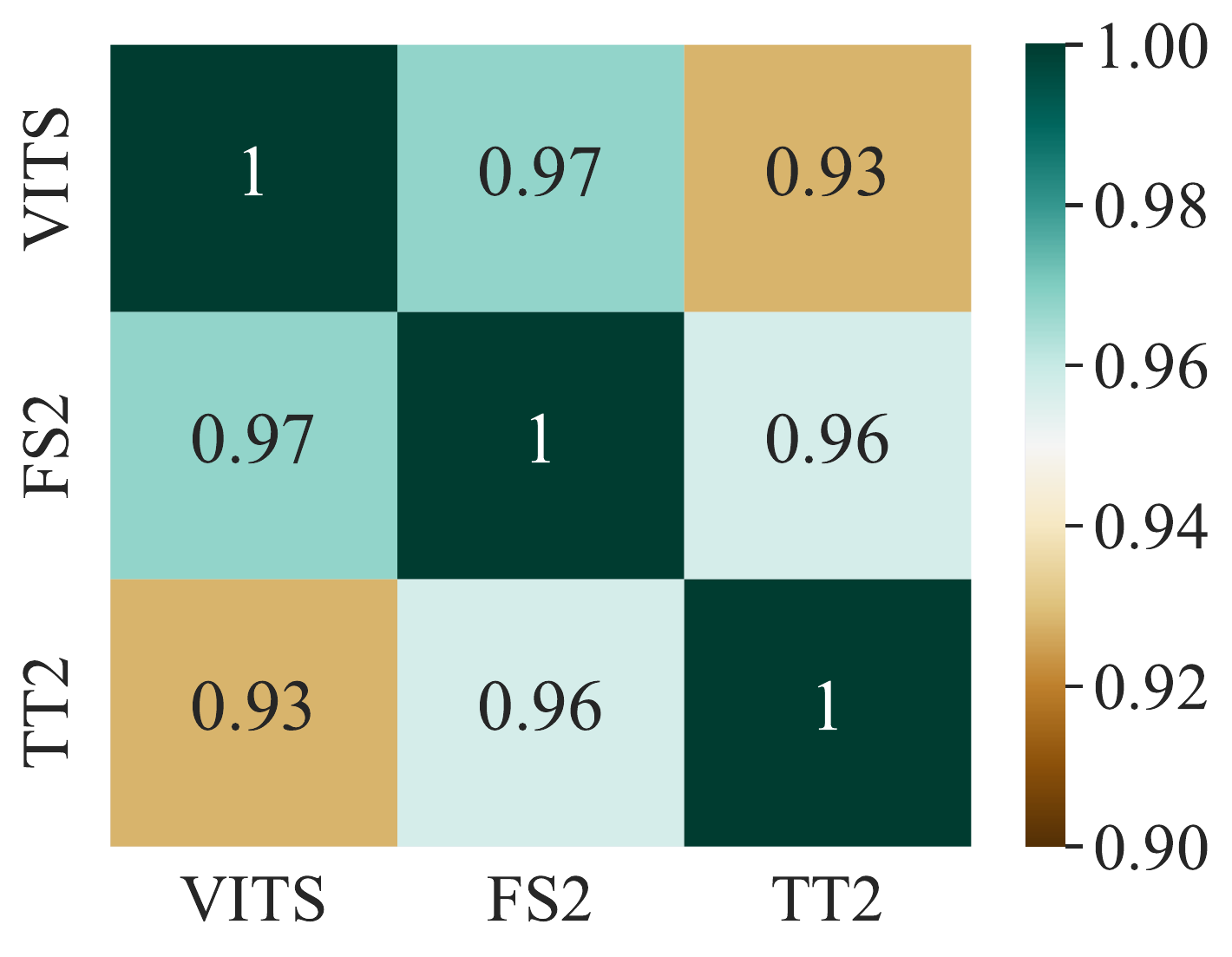}
    \vspace{-10pt}
  \caption{The Pearson correlation coefficients between different TTS systems: the unit distribution is collected with the development set of the Fisher Spanish-English dataset \cite{post2014fisher}. For the wave generation from FS2 and TT2, we both utilize the Hifi-GAN vocoder.}
  \label{fig:correlation}
  \vspace{-15pt}
\end{figure}

As in \cite{polyak2021speech, lakhotia2021generative}, speech discrete units from speech SSL representations can potentially disentangle linguistic, prosodic, and speaker-related information. However, at the same time, it is still noisy to use. For example, in \cite{lee2021textless}, the authors have shown that the same sentence spoken by different speakers, could result in different speech discrete unit sequences. A similar phenomenon may happen when the same sentence is generated by different TTS systems. To verify our hypothesis, we measure the Pearson correlation coefficients of the HuBERT units' distribution between different TTS systems trained on LJSpeech \cite{ito2017lj}.\footnote{The configuration of HuBERT and TTS systems are discussed in Sec.~\ref{ssec: experimental setings} in detail.} As shown in Fig.~\ref{fig:correlation}, it clearly indicates that different TTS systems are still different though with the same linguistic source and trained on the same corpus.  The data includes the development set of the Fisher Spanish-English corpus \cite{post2014fisher}.  On the other hand, given the same utterance in the target language, the synthesized speech should have the same linguistic content. Therefore, it is reasonable to assume that the extracted units could have a similar consensus shared across, given the same text is employed to generate speech from different TTS systems.

Following the assumption discussed above, we propose the framework as shown in Fig.~\ref{fig:framework}. The framework is based on the model proposed in \cite{lee2022direct} but is additionally designed to capture the high-level consensus over linguistic information across different TTS systems. To be specific, we add separate decoder branches for speech discrete units generated from different TTS systems. For simplicity, in Fig.~\ref{fig:framework}, we show the case with two targets, but it could be easily extended into three or more targets because of its parallel property. 

Instead of directly predicting the units in parallel, we also append a special token at the start of \textbf{target} unit sequences as shown in the gray blocks in Fig.~\ref{fig:framework}. The special token is defined as an indicator of the quality of synthesized speech, which we can use to select better output during model inference. Practically, we first compute the character error rates (CER) at the sentence level for each utterance from different TTS systems. Then, at the training stage, we assign token \textsc{[Y]} to the TTS system with the best CER among candidate target speech discrete units and token \textsc{[N]} to other systems.\footnote{For even cases, we assign \textsc{[Y]} to all systems with the best CER.} For inference, we compute the probability of the first predicted special token [Y] from all the decoders and select the one with the highest probability to continue generating the sequences. Noted that since the special token is at the start of the sequence, the inference process does not need to auto-regressively generate future tokens if it already has a lower probability than other branches. Therefore, compared to the base system without multiple TTS targets, there is not much additional searching burden when doing inference.

Due to the noise present in discrete units, a similar approach was explored in \cite{lee2021textless}. The authors proposed a speaker normalization method to normalize the units of different speakers to a reference speaker. However, in the case of using different TTS systems, it can be challenging to determine which system should be used as the reference, as they are all synthesized using the same text.

\section{Experiments}

\subsection{Datasets}
For S2ST, we use the Fisher Spanish-English dataset \cite{post2014fisher}, which is also widely used in the previous S2ST works \cite{lee2022direct, zhang2021uwspeech}. The English TTS systems are applied to synthesize target speech from the English text for training and validation. For the training of TTS systems, we use LJSpeech, a 24-hour single-speaker corpus \cite{ito2017lj}.

\subsection{Experimental Settings}
\label{ssec: experimental setings}

\subsubsection{Model architectures}

\noindent
\textbf{Speech-to-unit translation model (S2UT)}: We follow the updated version of S2UT model described in \cite{popuri2022enhanced}, which is an extension of \cite{lee2022direct}. For speech discrete unit generation, we adopt the pre-trained multilingual HuBERT, K-Means clustering, and the unit-based HFG vocoder released in \cite{lee2021textless, popuri2022enhanced}. The vocabulary size of the discrete unit is 1,000, corresponding to the number of clusters in the K-Means model. The generated discrete units are reduced by duplication-pooling, during the training of S2UT models. On the other hand, the reduced units are recovered to their original lengths through a duration predictor that is jointly trained with unit-based HFG. The same of \cite{popuri2022enhanced}, the encoder is initialized from a Conformer-based wav2vec 2.0, while the decoder is initialized from the mBART decoder \cite{liu2020multilingual} released in \cite{popuri2022enhanced}. To keep the consistency over different settings, we do not tune the hyper-parameters, but use the settings as \cite{popuri2022enhanced} for different systems.

\noindent
\textbf{TTS models}: As mentioned in Sec.~\ref{ssec: background}, we utilize three different TTS systems (i.e., TT2, FS2, and VITS) and three different vocoders (i.e., PWG, HFG, and SWG). To keep the reproducibility of the experiments, all the models are from public-available checkpoints in ESPnet-TTS \cite{hayashi2020espnet, hayashi2021espnet2}, an open-source framework for TTS.

\subsubsection{Training and decoding}

For the training of S2UT models, we use the AdamW optimizer \cite{loshchilov2018decoupled} with a learning rate of 0.0005. The scheduler is applied with a warmup policy that starts the learning rate from 1e-7 and reaches the maximum learning rate at 20k steps. We accumulate the gradients for every 120 steps to simulate a large batch size, which has shown to be effective for S2ST learning. During the training, we follow the ``LNA-D" policy introduced in \cite{popuri2022enhanced}, which does not fine-tune all the parameters in the pre-trained mBART decoder but only the LayerNorm and self-attention parameters. For the decoding of S2UT, we apply beam search with a beam size of 10.

For NAR TTS models, the duration can be tuned with a speed factor, resulting in the duration control to the generation. In our experiments, apart from the TTS models we discussed in Sec.~\ref{ssec: background}, we also apply different speed factors including 0.95, 1.0, and 1.05.

\subsubsection{Experiments design}
\label{sssec: experiment design}

The experiments generally include three folds: 

\noindent \textbf{Single TTS systems}: To compare the effect of different TTS systems, we directly train the S2UT model with discrete units generated from a single TTS system.

\noindent \textbf{Simple combination of TTS systems}: As discussed in Sec.~\ref{ssec: proposed framework}, multiple TTS systems could potentially improve the S2ST, by normalizing the noise from unit sequences. To systematically investigate the effects of different systems, we carry out experiments based on the modeling properties of different TTS systems, including AR versus NAR, different vocoders, different speed factors, and text2Mel versus text2wav.

\noindent \textbf{Multi-task framework for multiple TTS targets}: We follow the design of the experiments in the simple combination of TTS systems, but change from the data combination into the multi-task way of training, as introduced in Sec.~\ref{ssec: proposed framework}. We use the wav2vec 2.0-based ASR model to compute the CER mentioned in Sec.~\ref{ssec: proposed framework}. Due to the requirements of large GPU memory to train models with more than three additional branches, we limit the number of TTS systems to less than four.

\subsubsection{Evaluation metric}
\label{ssec: evaluation metric}

The TTS quality is first evaluated by inputting the synthesized speech to the ASR model and calculating the character error rate (CER) between the ASR prediction and the reference text. For the ASR model, we employ the open-source ASR model that is trained over wav2vec 2.0\footnote{\scriptsize{\url{https://huggingface.co/facebook/wav2vec2-large-960h-lv60-self}}}. For evaluation of the translation quality, we first utilize the same ASR model to get the transcription of the S2ST system (i.e., predicted units + code-HFG vocoder) and then compute BLEU score with the reference text using SacreBLEU \cite{post2018call}. Noted that the reference text is also tokenized and converted to lowercase without punctuation for BLEU calculation. 

\subsection{Results and Discussion}
\label{ssec: result}

\begin{table}
\centering
\caption{\label{tab: single TTS exp} S2ST Performances on different TTS systems. The ``Data ID" column stands for the target speech units generated from the TTS system. The CER and BLEU are calculated as discussed in Sec.~\ref{ssec: evaluation metric}. The acoustic models (AM) and vocoders are introduced in Sec.~\ref{ssec: background}.}
\begin{tabular}{l|l|c|cc}
\toprule 
Data ID & AM & Vocoder & CER($\downarrow$) & BLEU($\uparrow$) \\ 
\midrule
\textbf{A} &\multirow{3}{*}{TT2} & PWG & 9.1 & 37.3 \\
\textbf{B} & & HFG & 8.9 & 37.3 \\
\textbf{C} & & SWG & 8.7 & 37.3 \\
\midrule
Avg. & TT2 & / & 8.9 & 37.3 \\
\midrule
\textbf{D} & \multirow{3}{*}{FS2} & PWG & 9.4 & 37.5\\
\textbf{E} & & HFG & \textbf{8.3} & 37.6 \\
\textbf{F} & & SWG & 8.7 & 37.5 \\
\midrule
Avg. & FS2 & / & 8.8 & 37.5 \\
\midrule
\textbf{G} & \multicolumn{2}{l|}{VITS} & 8.4 & \textbf{38.3}\\
\bottomrule
\end{tabular}

\end{table}


We conduct experiments with target speech discrete units generated from a single TTS system. The best system was obtained from synthesized speech using VITS (data \textbf{G}), and the results are shown in Table~\ref{tab: single TTS exp}. We observe that there was no significant effect on S2ST performance when different vocoders were applied to models \textbf{A}-\textbf{C} and models \textbf{D}-\textbf{F}. However, the difference in acoustic models could affect the S2ST results, with the best system using target speech synthesized from VITS and the worst from TT2. We also find that the CER in each TTS acoustic model roughly correlated with their S2ST performance.

An interesting finding from Table~\ref{tab: single TTS exp} is that systems with different vocoders achieved similar performances despite having different CERs from the ASR model. We assume that this is due to the HuBERT units helping to normalize the vocoder differences in speech synthesis. To verify this hypothesis, we conduct a Pearson coefficient analysis over HuBERT unit distribution on FS2 with different vocoders and found that the Pearson scores were all above 0.98.

Furthermore, we find that VITS usually had a higher CER than TT2 and FS2 for spoken words (e.g., "HMM", "HUM"). This could be due to the data domain mismatch between the read speech used for TTS training (i.e., LJSpeech) and ASR training (i.e., Librispeech) and the conversational speech used for S2ST training (i.e., Fisher).


Table~\ref{tab: SF exp} presents the effect of different speed factors on NAR TTS systems. The results show that a smaller speed factor can improve the performance of the S2ST system. However, when measuring the CER of the synthesized speech, using the default speed factor of 1.0 is more favorable. Combining the results from Table~\ref{tab: single TTS exp} and Table~\ref{tab: SF exp}, we find that VITS with a speed factor of 0.95 yields the best-synthesized data for building the unit-based S2ST system. Therefore, we report this number as a reference in the following combination experiments.

\begin{table}
\centering
\caption{\label{tab: SF exp} S2ST Performances on different speed factors. The Fastspeech2 (FS) model is combined with the HFG vocoder for TTS as default. The TTS models are introduced in Sec.~\ref{ssec: background}.}
\begin{tabular}{l|l|c|cc}
\toprule 
Data ID & TTS & Speed Factor & CER($\downarrow$) & BLEU($\uparrow$) \\ 
\midrule
\textbf{H} & \multirow{3}{*}{FS2} & 0.95 & 8.7 & 38.0\\
\textbf{E} & & 1.0 & 8.3 & 37.5\\
\textbf{I} & & 1.05 & 9.5 & 37.4\\
\midrule
\textbf{J} & \multirow{3}{*}{VITS} & 0.95 & 8.6 & \textbf{38.7} \\
\textbf{G} & & 1.0 & 8.4 & 38.3\\
\textbf{K} & & 1.05 & 8.5 & 38.3\\
\bottomrule
\end{tabular}
\end{table}

\begin{table}
\centering
\caption{\label{tab: combine tts} S2ST Performances on multiple TTS targets. We follow the categories listed in Sec.~\ref{sssec: experiment design} to conduct experiments The models with \cmark in the ``Multi-task" column are trained with the framework proposed in Sec.~\ref{ssec: proposed framework}.}
\begin{tabular}{l|l|c|c}
\toprule 
Category & Data & Multi-task & BLEU($\uparrow$) \\ 
\midrule
Best Single TTS system & \textbf{J} & / & 38.7 \\
\midrule
\multirow{2}{*}{TT2 + FS2} & \multirow{2}{*}{\textbf{B} + \textbf{E}} & \xmark & 37.0\\
& & \cmark & 37.6 \\
\midrule
\multirow{2}{*}{TT2 + Vocoders} & \multirow{2}{*}{\textbf{A} + \textbf{B} + \textbf{C}} & \xmark & 37.3\\
& & \cmark & 37.3\\
\midrule
\multirow{2}{*}{FS2 + Vocoders} & \multirow{2}{*}{\textbf{D} + \textbf{E} + \textbf{F}} & \xmark & 37.7 \\
& & \cmark & 37.9 \\
\midrule
\multirow{2}{*}{FS2 + Speed Factors} &  \multirow{2}{*}{\textbf{H} + \textbf{E} + \textbf{I}} & \xmark & 38.8 \\
& & \cmark & 39.7 \\
\midrule
\multirow{2}{*}{VITS + Speed Factors} & \multirow{2}{*}{\textbf{J} + \textbf{G} + \textbf{K}} & \xmark & 39.7\\
& & \cmark & \textbf{41.5}\\
\midrule
\multirow{2}{*}{VITS + TT2} &  \multirow{2}{*}{\textbf{B} + \textbf{G}} & \xmark & 37.2 \\
& & \cmark & 38.4\\
\midrule
\multirow{2}{*}{VITS + FS2} & \multirow{2}{*}{\textbf{E} + \textbf{G}} & \xmark & 39.6 \\
& & \cmark & 40.5 \\
\midrule
\multirow{2}{*}{VITS + TT2 + FS2} & \multirow{2}{*}{\textbf{B} + \textbf{E} + \textbf{G}} & \xmark & 38.5\\
& & \cmark & 40.1\\
\bottomrule
\end{tabular}

\end{table}

\begin{table}[]
    \centering
\caption{S2ST Performances from different branches of the proposed framework. The BLEU Diff. is the absolute difference between the system trained on corresponding single TTS systems. The details are discussed in Sec.~\ref{ssec: result}.}
    \begin{tabular}{l|l|cc}
    \toprule
       Category & Branch(es) & BLEU($\uparrow$) & BLEU Diff.\\
       \midrule
        \multirow{4}{*}{VITS + TT2 + FS2} & \textbf{B} & 37.9 & +0.6\\
        & \textbf{E} & 38.8 & +1.2\\
        & \textbf{G} & 38.9 & +0.6\\
        & \textbf{B} + \textbf{E} + \textbf{G} & 40.1 & / \\
        \midrule
        \multirow{3}{*}{VITS + FS2} & \textbf{E} & 38.8 & +1.2\\
        & \textbf{G} & 39.0 & +0.7 \\
        & \textbf{E} + \textbf{G} & 40.5 & / \\
        \midrule
        \multirow{4}{*}{VITS + Speed Factors} & \textbf{J} & 39.5 & +0.8 \\
        & \textbf{G} & 39.1 & +0.8 \\
        & \textbf{K} & 39.0 & +0.7 \\
        & \textbf{J} + \textbf{G} + \textbf{K} & 41.5 & / \\
    \bottomrule
    \end{tabular}
    \label{tab:abla study}
\end{table}

Table~\ref{tab: combine tts} shows the results with the combination of different TTS systems \textbf{A}-\textbf{K}, investigated in previous experiments. The experiments with ``\xmark" are the approaches that simply combine the data from different TTS systems for training, while the experiments with ``\cmark" are based on our proposed method in Sec.~\ref{ssec: proposed framework}. When comparing models between Table~\ref{tab: single TTS exp} and Table~\ref{tab: combine tts}, the results show that there are usually some improvements in S2ST by simply merging the data from different TTS systems. For example, we get 39.6 BLEU by combining data \textbf{E} and \textbf{G} from Table~\ref{tab: combine tts}, while from Table~\ref{tab: single TTS exp}, when training with only \textbf{B} or \textbf{G}, we get 37.6 and 38.3 BLEU, respectively. Noted that there are also cases that the simple combination of data does not improve the S2ST performances (e.g., \textbf{A} + \textbf{B} + \textbf{C} versus \textbf{A}, \textbf{B}, and \textbf{C}). Meanwhile, compared to models without multi-targets training, we would get even better results by adopting the framework proposed in Sec.~\ref{ssec: proposed framework}. 

As introduced in Sec.~\ref{ssec: proposed framework}, given an utterance during inference, the proposed framework starts with predicting the first token (e.g., [Y] or [N]) of each decoder branch. Then, it selects the decoder branch with the highest probability of token [Y] as the decoder branch for this utterance. In Table~\ref{tab:abla study}, we conduct an ablation study about the inference based on the special token. To be specific, we report the S2ST performances of each decoder branch from the proposed framework  and compare the results with S2ST systems trained only on the corresponding TTS targets. Three S2ST systems with the top three performances are chosen in our comparison. The results show that the S2ST performances get improved for all the branches when compared to systems trained on a single TTS system (e.g., models trained with \textbf{E} or \textbf{G} from Table~\ref{tab: single TTS exp}, and models trained with \textbf{J}, \textbf{G}, \textbf{K} from Table~\ref{tab: SF exp}). Meanwhile, it also shows the effectiveness of the proposed inference procedure because the combination of branches with our proposed method still outperforms the single-branch performances.


\section{Conclusion}
This work first investigates the effect of using different targets from different TTS systems. Experiments show that simply combining the target speech from TTS systems could help the learning of S2ST, especially the speech-to-unit model (S2UT). Following the findings, we propose a new framework to integrate multiple TTS targets into the S2ST modeling. Experiments demonstrate that our proposed framework can consistently improve the performances of the best baseline S2ST by 2.8 BLEU.

\section{Acknowledgement}
This work was supported by a Meta AI SRA grant. Jiatong Shi and Shinji Watanabe are funded in part of the Bridges system ~\cite{nystrom2015bridges}, which is supported by NSF award number ACI-1445606, at the Pittsburgh Supercomputing Center (PSC).



\ninept
\bibliographystyle{IEEEbib}
\bibliography{strings,refs}

\begin{thebibliography}{10}

\bibitem{vidal1997finite}
Enrique Vidal,
\newblock ``Finite-state speech-to-speech translation,''
\newblock in {\em ICASSP}, 1997.

\bibitem{matsuda2013multilingual}
Shigeki Matsuda, Xinhui Hu, Yoshinori Shiga, et~al.,
\newblock ``Multilingual speech-to-speech translation system: Voicetra,''
\newblock in {\em ICMDM}, 2013.

\bibitem{do2016preserving}
Quoc~Truong Do, Tomoki Toda, Graham Neubig, , et~al.,
\newblock ``Preserving word-level emphasis in speech-to-speech translation,''
\newblock {\em TASLP}, 2016.

\bibitem{jia2019direct}
Ye~Jia, Ron~J Weiss, Fadi Biadsy, et~al.,
\newblock ``Direct speech-to-speech translation with a sequence-to-sequence
  model,''
\newblock {\em Interspeech}, 2019.

\bibitem{jia2022translatotron}
Ye~Jia, Michelle~Tadmor Ramanovich, Tal Remez, , et~al.,
\newblock ``Translatotron 2: High-quality direct speech-to-speech translation
  with voice preservation,''
\newblock in {\em ICML}, 2022.

\bibitem{jia2022leveraging}
Ye~Jia, Yifan Ding, Ankur Bapna, et~al.,
\newblock ``{Leveraging unsupervised and weakly-supervised data to improve
  direct speech-to-speech translation},''
\newblock in {\em Interspeech}, 2022, pp. 1721--1725.

\bibitem{lee2022direct}
Ann Lee, Peng-Jen Chen, Changhan Wang, et~al.,
\newblock ``Direct speech-to-speech translation with discrete units,''
\newblock in {\em ACL}, 2022.

\bibitem{kano2021transformer}
Takatomo Kano, Sakriani Sakti, and Satoshi Nakamura,
\newblock ``Transformer-based direct speech-to-speech translation with
  transcoder,''
\newblock in {\em SLT}, 2021.

\bibitem{zhang2021uwspeech}
Chen Zhang, Xu~Tan, Yi~Ren, et~al.,
\newblock ``Uwspeech: Speech to speech translation for unwritten languages,''
\newblock in {\em AAAI}, 2021.

\bibitem{ma2021direct}
Xutai Ma, Hongyu Gong, Danni Liu, et~al.,
\newblock ``Direct simultaneous speech to speech translation,''
\newblock {\em arXiv preprint arXiv:2110.08250}, 2021.

\bibitem{popuri2022enhanced}
Sravya Popuri, Peng-Jen Chen, Changhan Wang, et~al.,
\newblock ``{Enhanced Direct Speech-to-Speech Translation Using Self-supervised
  Pre-training and Data Augmentation},''
\newblock in {\em Interspeech}, 2022, pp. 5195--5199.

\bibitem{lee2021textless}
Ann Lee, Hongyu Gong, Paul-Ambroise Duquenne, et~al.,
\newblock ``Textless speech-to-speech translation on real data,''
\newblock in {\em NAACL}, 2022, pp. 860--872.

\bibitem{kikui2003creating}
Genichiro Kikui, Eiichiro Sumita, Toshiyuki Takezawa, et~al.,
\newblock ``Creating corpora for speech-to-speech translation,''
\newblock in {\em Eurospeech}, 2003.

\bibitem{jia2022cvss}
Ye~Jia, Michelle~Tadmor Ramanovich, et~al.,
\newblock ``{CVSS} corpus and massively multilingual speech-to-speech
  translation,''
\newblock in {\em LREC}, 2022, pp. 6691--6703.

\bibitem{jeuris2022libris2s}
Pedro Jeuris and Jan Niehues,
\newblock ``Libris2s: A german-english speech-to-speech translation corpus,''
\newblock in {\em LREC}, 2022, pp. 928--935.

\bibitem{shen2020non}
Jonathan Shen, Ye~Jia, Mike Chrzanowski, et~al.,
\newblock ``Non-attentive {T}acotron: Robust and controllable neural {TTS}
  synthesis including unsupervised duration modeling,''
\newblock {\em arXiv preprint arXiv:2010.04301}, 2020.

\bibitem{ren2020fastspeech}
Yi~Ren, Chenxu Hu, Xu~Tan, et~al.,
\newblock ``Fastspeech 2: Fast and high-quality end-to-end text to speech,''
\newblock in {\em ICLR}, 2020.

\bibitem{post2014fisher}
Matt Post, Gaurav Kumar, Adam Lopez, et~al.,
\newblock ``Fisher and {CALLHOME} spanish--english speech translation,''
\newblock {\em LDC2014T23. Web Download. Philadelphia: Linguistic Data
  Consortium}, 2014.

\bibitem{yang2021superb}
Shu-wen Yang, Po-Han Chi, Yung-Sung Chuang, et~al.,
\newblock ``{SUPERB}: Speech processing universal performance benchmark,''
\newblock in {\em Proc. Interspeech 2021}, 2021, pp. 1194--1198.

\bibitem{tsai2022superb}
Hsiang-Sheng Tsai, Heng-Jui Chang, Wen-Chin Huang, et~al.,
\newblock ``{SUPERB-SG}: Enhanced speech processing universal performance
  benchmark for semantic and generative capabilities,''
\newblock in {\em ACL}, 2022, pp. 8479--8492.

\bibitem{polyak2021speech}
Adam Polyak, Yossi Adi, Jade Copet, et~al.,
\newblock ``Speech resynthesis from discrete disentangled self-supervised
  representations,''
\newblock in {\em Interspeech}, 2021.

\bibitem{huang2022s3prl}
Wen-Chin Huang, Shu-Wen Yang, Tomoki Hayashi, et~al.,
\newblock ``S3{PRL-VC}: Open-source voice conversion framework with
  self-supervised speech representations,''
\newblock in {\em ICASSP}, 2022.

\bibitem{hayashi2020discretalk}
Tomoki Hayashi and Shinji Watanabe,
\newblock ``Discretalk: Text-to-speech as a machine translation problem,''
\newblock {\em arXiv preprint arXiv:2005.05525}, 2020.

\bibitem{lakhotia2021generative}
Kushal Lakhotia, Eugene Kharitonov, Wei-Ning Hsu, et~al.,
\newblock ``On generative spoken language modeling from raw audio,''
\newblock {\em TACL}, 2021.

\bibitem{aguero2006prosody}
PD~Aguero, Jordi Adell, and Antonio Bonafonte,
\newblock ``Prosody generation for speech-to-speech translation,''
\newblock in {\em ICASSP}, 2006.

\bibitem{hsu2021hubert}
Wei-Ning Hsu, Benjamin Bolte, Yao-Hung~Hubert Tsai, et~al.,
\newblock ``Hu{BERT}: Self-supervised speech representation learning by masked
  prediction of hidden units,''
\newblock {\em TASLP}, 2021.

\bibitem{shen2018natural}
Jonathan Shen, Ruoming Pang, Ron~J Weiss, et~al.,
\newblock ``Natural {TTS} synthesis by conditioning {W}avenet on {M}el
  spectrogram predictions,''
\newblock in {\em ICASSP}, 2018.

\bibitem{kim2021conditional}
Jaehyeon Kim, Jungil Kong, and Juhee Son,
\newblock ``Conditional variational autoencoder with adversarial learning for
  end-to-end text-to-speech,''
\newblock in {\em ICML}, 2021.

\bibitem{yamamoto2020parallel}
Ryuichi Yamamoto, Eunwoo Song, and Jae-Min Kim,
\newblock ``Parallel {W}ave{GAN}: A fast waveform generation model based on
  generative adversarial networks with multi-resolution spectrogram,''
\newblock in {\em ICASSP}, 2020.

\bibitem{kong2020hifi}
Jungil Kong, Jaehyeon Kim, and Jaekyoung Bae,
\newblock ``Hi{F}i-{GAN}: Generative adversarial networks for efficient and
  high fidelity speech synthesis,''
\newblock {\em NeurIPS}, 2020.

\bibitem{mustafa2021stylemelgan}
Ahmed Mustafa, Nicola Pia, and Guillaume Fuchs,
\newblock ``Stylemelgan: An efficient high-fidelity adversarial vocoder with
  temporal adaptive normalization,''
\newblock in {\em ICASSP}, 2021.

\bibitem{ito2017lj}
Keith Ito and Linda Johnson,
\newblock ``The {LJ} speech dataset,'' 2017.

\bibitem{liu2020multilingual}
Yinhan Liu, Jiatao Gu, Naman Goyal, et~al.,
\newblock ``Multilingual denoising pre-training for neural machine
  translation,''
\newblock {\em TACL}, 2020.

\bibitem{hayashi2020espnet}
Tomoki Hayashi, Ryuichi Yamamoto, Katsuki Inoue, et~al.,
\newblock ``{ESP}net-{TTS}: Unified, reproducible, and integratable open source
  end-to-end text-to-speech toolkit,''
\newblock in {\em ICASSP}, 2020.

\bibitem{hayashi2021espnet2}
Tomoki Hayashi, Ryuichi Yamamoto, Takenori Yoshimura, et~al.,
\newblock ``E{SP}net2-{TTS}: Extending the edge of {TTS} research,''
\newblock {\em arXiv preprint arXiv:2110.07840}, 2021.

\bibitem{loshchilov2018decoupled}
Ilya Loshchilov and Frank Hutter,
\newblock ``Decoupled weight decay regularization,''
\newblock in {\em ICLR}, 2018.

\bibitem{post2018call}
Matt Post,
\newblock ``A call for clarity in reporting {BLEU} scores,''
\newblock in {\em Conference on Machine Translation}, 2018.

\bibitem{nystrom2015bridges}
Nicholas~A Nystrom, Michael~J Levine, Ralph~Z Roskies, et~al.,
\newblock ``Bridges: a uniquely flexible hpc resource for new communities and
  data analytics,''
\newblock in {\em XSEDE}, 2015.

\end{thebibliography}

\end{document}